\documentclass[11pt,a4paper]{article}
\usepackage{jheppub}

\usepackage{amssymb}
\usepackage{amsmath}
\usepackage{epsfig}
\usepackage{amssymb}
\usepackage{epsf}
\usepackage{epsfig}
\usepackage{color}
\usepackage{ifpdf}
\usepackage{cite}
\usepackage{cases}
\usepackage{subfigure}
\usepackage{multirow}

\newcommand{\ba}{\begin{eqnarray}}
\newcommand{\ea}{\end{eqnarray}}
\newcommand{\be}{\begin{equation}}
\newcommand{\ee}{\end{equation}}
\newcommand{\bd}{\begin{displaymath}}
\newcommand{\ed}{\end{displaymath}}
\newcommand{\een}{\nonumber\end{equation}}
\newcommand{\bea}{\begin{eqnarray}}
\newcommand{\eean}{\nonumber\end{eqnarray}}
\newcommand{\eea}{\end{eqnarray}}

\newcommand{\xoct}{\langle x \rangle^{(8)}}
\newcommand{\xiso}{\langle x \rangle^{(3)}}

\def\eq#1{Eq.~(\ref{#1})}
\def\eqs#1{Eqs.~(\ref{#1})}
\def\fig#1{Fig.~\ref{#1}}

\def\tab#1{Table~\ref{#1}}

\def\cite#1{\citep{#1}}

\newcommand{\gap}{\hspace{10pt}}

\newcommand{\mev}{\mathrm{MeV}}
\newcommand{\gev}{\mathrm{GeV}}
\newcommand{\fm}{\mathrm{fm}}

\newcommand{\mps}{m_{\rm{PS}}}

\def\mcC{{\mathcal C}}
\def\mcD{{\mathcal D}}
\def\mcJ{{\mathcal J}}

\def\mcO{{\mathcal O}}

\def\la{\langle}
\def\ra{\rangle}
\def\psibar{\overline{\psi}}
\def\chibar{\overline{\chi}}

\newcommand{\old}[1]{}

\newcommand{\plotangle}{0}

\title{First moment of the flavour octet nucleon parton distribution function using lattice QCD}

\collaborationImg{\includegraphics{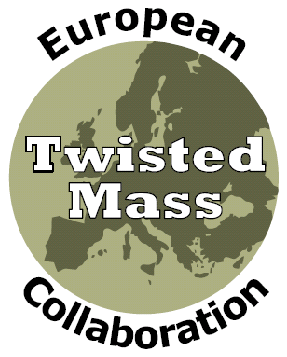}}
\author[a,b]{Constantia Alexandrou}
\affiliation[a]{Department of Physics, University of Cyprus, P.O. Box 20537, 1678 Nicosia, Cyprus}
\affiliation[b]{Computation-based Science and Technology Research Center (CaSToRC), The Cyprus Institute,20 Constantinou Kavafi Street Nicosia 2121, Cyprus}

\author[a]{Martha Constantinou}

\author[c]{Simon Dinter}
\affiliation[c]{NIC, DESY, Platanenallee 6, D-15738 Zeuthen, Germany}

\author[d]{Vincent Drach}
\author[a]{Kyriakos Hadjiyiannakou}
\affiliation[d]{$CP^3$-Origins \& the  Danish Institute for Advanced Study  DIAS, University of Southern Denmark, Campusvej 55, DK-5230 Odense M, Denmark}

\author[a,b,c]{Karl Jansen}
\author[a]{Giannis Koutsou}
\author[e]{Alejandro Vaquero}
\affiliation[e]{INFN, Sezione di Milano-Bicocca
Edificio U2, Piazza della Scienza 3
20126 Milano, Italy}

\abstract{We perform a lattice computation of
the flavour octet contribution to the average quark momentum
in a nucleon, $\xoct_{\mu^2 = 4~\gev^2 }$. In particular, we fully take the
disconnected contributions into account in our analysis for which
we use
a generalization of the technique developed in \cite{Dinter:2012tt}.
We investigate systematic effects with a particular emphasis on
the excited states contamination. We find that in
the renormalization free ratio
$\frac{\la x \ra^{(3)}}{\la x \ra^{(8)}}$ (with $\la x \ra^{(3)}$ the
non-singlet moment) the excited state contributions cancel to a
large extend making this ratio a promising candidate for a comparison
to phenomenological analyses.
Our final result for this ratio is in agreement with the phenomenological
value and we
find, including systematic errors, $\frac{\la x \ra^{(3)}}{\la x \ra^{(8)}} = 0.39(1)(4)$.}

\keywords{parton distribution function,lattice QCD}
\emailAdd{drach@cp3.dias.sdu.dk}
\emailAdd{karl.jansen@desy.de}
\preprint{DESY 15-008,CP3-Origins-2015-004 DNRF90, DIAS-2015-4}

\begin{document}

\maketitle

\section{Introduction}

Computing nucleon properties from first principles using lattice QCD is a long
standing challenge. In particular, moments of parton distribution functions (pdfs)
are important benchmark quantities for lattice calculations and provide insight
into the structure of hadrons.
The computation of various moments of pdfs is therefore a very active
research area for lattice calculations, see
\cite{Alexandrou:2013joa,Alexandrou:2011nr,Alexandrou:2011db,Alexandrou:2010hf,Bali:2013dpa,Bali:2012av,Green:2012ud,Hagler:2007xi,Alexandrou:2013cda,Constantinou:2014tga} for recent results.
In the work we perform here a
first lattice calculation of the octet contribution to
the average quark momentum in a nucleon, $\xoct_{\mu^2}$ is
presented, where $\mu^2$ denotes the renormalization scale.
This quantity is of  interest by itself and can be extracted
from phenomenological analyses of data from deep inelastic scattering
experiments, see below.
It therefore can serve as an additional quantity to probe QCD and
the structure of hadrons, see \cite{Constantinou:2014tga}.

In addition, $\xoct_{\mu^2}$ needs the same renormalization
constant as the iso-vector averaged quark momentum in the nucleon $\xiso_{\mu^2}$ and therefore
in the ratio of $\xoct_{\mu^2}$ and $\xiso_{\mu^2}$ the renormalization
constant cancels.  Although we can evaluate non-perturbatively the
renormalization constant, its cancelation in the ratio eliminates any
uncertainty related to its determination. Comparing the ratio to a
phenomenological analysis can thus help to understand, whether
renormalization effects can play a role in the presently observed
discrepancy between lattice calculations of
$\xiso_{\mu^2}$ and phenomenological determinations of this quantity.

Despite the above given motivations to compute $\xoct_{\mu^2}$ there
is so far no value of this quantity available from a lattice
determination. The reason for this is that $\xoct_{\mu^2}$ is very
difficult to calculate since it involves dis-connected, singlet contributions.
The definition of the flavour octet moment $\xoct_{\mu^2}$ reads :
\be\label{def:x8}
\la x \ra^{(8)}_{\mu^2} = \int_{-1}^1 dx~x \left[ u(x,\mu^2) + d(x,\mu^2) - 2 s(x,\mu^2)) \right]
\ee
where $q(x,\mu)$ denotes the sum of the parton distribution function of the quark $q$ and anti-quark  $\bar{q}$.
Using the same convention, we also define the non-singlet
contribution $\la x \ra^{(3)}_{\mu^2}$,
\be\label{def:x3}
\la x \ra^{(3)}_{\mu^2} = \int_{-1}^1 dx~x \left[ u(x,\mu^2) - d(x,\mu^2)  \right]\; .
\ee

At a value of the renormalization scale of $\mu^2=4{\rm GeV}^2$
the two moments$\la x \ra^{(3)}_{\mu^2 = 4 ~\gev^2} $  and
$\la x \ra^{(8)}_{\mu^2 = 4 ~\gev^2} $ can be extracted phenomenologically using
parton distribution functions determined from deep inelastic
scattering data, and read, using the ABM12 pdfs
set \cite{Alekhin:2013nda} and the analysis of ref.~\cite{Sergey}

\be\label{eq:pheno_x}
\la x \ra^{(3)}_{\mu^2 = 4 ~\gev^2} = 0.153(4),\gap \la x \ra ^{(8)}_{\mu^2 = 4 ~\gev^2} = 0.470(7).
\ee

The quantities $\la x \ra^{(3)}_{\mu^2}$ and $\la x \ra^{(8)}_{\mu^2}$ are related to
matrix elements of local operators that
can be computed in Euclidean space-time and are hence accessible to lattice QCD
calculations. Introducing

\be\label{eq:twist2_unpol}
O^a_{\{\mu_1\cdots\mu_n\}} =\psibar \gamma_{\{\mu_1} i \overleftrightarrow{D}_{\mu_2}\cdots i \overleftrightarrow{D}_{\mu_n\}} \lambda^a \psi \\
\ee
where $\lambda^{a}$ are the Gell-Mann matrices acting on a three flavour
quark field $\psi=(u,d,s)$. With $\overleftrightarrow{D}$ we denote the
symmetrized covariant lattice derivative and with the curly brackets
the symmetrization and subtraction of the trace. The required matrix element
is then given by

\be\label{eq:rel_Oa_xa}
\la N(p,s) |O^{a=3,8}_{\{\mu\nu\}}|  N(p,s) \ra\lvert_{\mu^2} =
\la x \ra^{(3,8)}_{\mu^2}  \bar{u}_N(p,s) \gamma_{\{\mu} P_{\nu\}} u_N(p,s)\; .
\ee

From the definition in eq.~(\ref{def:x8}) it is clear that the octet matrix element
$\la N(p,s) |O^{a=8}_{\{\mu\nu\}}|  N(p,s) \ra$ involves
dis-connected contributions
which have in
general a bad signal to noise ratio, require thus a very high statistics and
are consequently very difficult to compute on the lattice.
In addition, the dis-connected contribution to $\la x \ra^{(8)}_{\mu^2}$
is an
$SU(3)_{\rm flavour}$ breaking effect and is thus expected to be small.
In \cite{Dinter:2012tt} and \cite{Alexandrou:2013wca} we have demonstrated
that for the here used (maximally) twisted mass lattice discretization of QCD
there are special noise reduction techniques which can help substantially
to improve on the signal to noise ratio. Indeed, these techniques allowed
us to compute a number of quantities which were very difficult
to access before \cite{Alexandrou:2013nda,Abdel-Rehim:2013wlz}.

In this work we present a generalization of the particular technique used in the
context of the determination of the $\sigma$-terms \cite{Dinter:2012tt}
to calculate non-perturbatively the disconnected contribution relevant
for $\la x \ra^{(8)}_{\mu^2}$. As we will see below, this generalization
will indeed provide a statistically significant signal for $\la x \ra^{(8)}_{\mu^2}$.
Note that  contrary to the
case of the $\sigma$-terms, the formalism to compute
dis-connected contributions developed here is not limited to the twisted
mass formulation. The technique could also be applied to other flavour
octet operators which would allow to determine non perturbatively
moments of polarized pdfs or the flavour octet axial
coupling of the nucleon $g_A^{(8)}$ for instance, see e.g.~\cite{Alexandrou:2013wca,Abdel-Rehim:2013wlz}.

The paper is organized as follow : after describing the basic ingredients
of our computation we give the details of the variance reduction technique used
in this work. We then present a study of the systematic effects appearing in our
calculation and
perform an extrapolation of the results to the physical pion mass in order to
compare with the phenomenologically obtained values.

\section{ Simulation Details}

The lattice action used in our simulations includes
as active degrees of freedom,
besides the gluon field,
a mass-degenerate light up and down
quark doublet as well as a strange-charm quark pair in the sea, a situation which
we refer to as the $N_f=2+1+1$ setup.
We use the Iwasaki action~\cite{Iwasaki:1985we} for the pure gauge
action and the twisted mass fermion formulation for the Dirac action.
In particular, we make use of the formulation of
refs.~ \cite{Frezzotti:2000nk,Frezzotti:2003ni} for the light mass degenerate
$u$--$d$ sector, while the action introduced in
refs.~\cite{Frezzotti:2003xj,Frezzotti:2004wz} is employed
for the mass non-degenerate $c$--$s$ sector.
The quark mass parameters of the heavy flavour pair have
been tuned so that in the unitary lattice setup the Kaon and D-meson masses,
take approximately their
experimental values. More information about the $N_f=2+1+1$ setup scheme and further simulation
details can be found in ref.~\cite{Baron:2010bv,Baron:2010th}.
For the results we will show here, we have employed
two values of the lattice spacing determined using the nucleon mass in
\cite{Alexandrou:2013joa}. They read $a\approx 0.082~\fm$ ($\beta=1.95$)
and $a\approx 0.064~\fm$ ($\beta=2.1$).
In addition, we will use a number of quark masses corresponding to
pion masses in the range of $300~\mev - 500~\mev$.  The
parameters of the ensembles used in this work are summarized in \tab{tab:runs}.

\begin{table}[h]
\begin{center}
\begin{tabular}{c|c|c|c|c}
label & $\beta$  & $a\,\mu_l$   & Volume &  $\mps~[MeV] $ \\
\hline
B35.32     & 1.95  &    0.0035 &  32$^3\times$64 & 302 \\
B55.32     & 1.95  &    0.0055 &  32$^3\times$64& 372 \\
B75.32     & 1.95  &    0.0075  &  32$^3\times$64& 432 \\
B85.24     & 1.95  &    0.0085  &  24$^3\times$48& 466  \\
\hline
D45.32sc &  2.1 &    0.0045  &  32$^3\times$64 & 372  \\
\end{tabular}
\caption{Ensembles used in this work and relevant parameters.}
\label{tab:runs}
\end{center}
\end{table}

In order to fix the notation, we introduce the twisted mass lattice Dirac operator
$D_{f,\rm tm}$
for a doublet of mass degenerate quarks :
\be
D_{f,\rm tm}[U]  = D_{\rm W}[U] + ia\mu_f\gamma_5 \tau^3 \; .
\label{eq:Dtm_nf2}
\ee
Here  $D_{\rm W}[U]$ is the Wilson Dirac operator, $\mu_f$ denotes the
bare twisted mass and $\tau^3$ is the third Pauli matrix.
For further needs we also introduce the operators $D_{f,\pm}$ denoting the
upper and lower flavour components of  $D_{f,\rm tm}[U] $,
referred to as the Osterwalder-Seiler (OS) lattice Dirac operator:
\be
D_{f,\pm}[U] = \rm{tr}~\left\{\frac{1\pm\tau_3}{2}D_{f,\rm tm}[U] \right\},
\label{eq:Dpm}
\ee
where $\rm{tr}$ denotes the trace in flavour space.
We will call
$D_{f,\pm}[U]$ the lattice Dirac operator of an Osterwalder-Seiler quark with mass $\pm \mu_f$.

When we discuss below the 2-point and 3-point correlation functions necessary for this work,
we will use the so-called physical basis of quark fields denoted as $\psi_f$.
The physical field basis is related to the twisted quark field basis, $\chi_f$, by
the following field rotation,

\be\label{eq:rotation_phys}
\psi_f \equiv e^{i\frac{\omega_f}{2} \gamma_5 \tau^3} \chi_f \gap\textmd{and}\gap  \psibar_f \equiv \chibar_f  e^{i\frac{\omega_f}{2} \gamma_5 \tau^3},
\ee
where the twist angle $\omega_f=\pi/2$ at maximal twist.
In addition, $\psi_f$ with index $f=l,s$ will denote  quark field
doublets  of light ($l$) or  strange ($s$) quarks depending on
the mass $\mu_f$ chosen in the valence sector. Since $\psi_f$ will always
refer to the physical basis we will denote  with $u$ and $d$ the
two components of $\psi_l$. Following the notation
of \eq{eq:Dpm} we will denote with $s_\pm$ the two components of $\psi_s$.
Employing the OS Dirac operator in the valence sector for the strange
quark leads to a
mixed action where the strange OS quark mass has been tuned to match
within errors the unitary Kaon mass.

\subsection{Nucleon matrix elements}

The nucleon two-point function is defined in the physical basis by
\be\label{eq:C2pts}
C^{\pm }_{N,\rm 2pts} (t) =   \sum_{\vec{x}}   \rm{tr}~\Gamma^{\pm}  \la \mcJ_{N}(x) \overline{\mcJ}_{N}(0)  \ra ,
\ee
where  the source position is fixed to 0 in order to lighten notations and
$t$ thus denotes the source-sink separation. We also introduced the
parity projectors  $\Gamma^{\pm} = (1\pm\gamma_0)/{2}$. The subscript
$N$ refers to the proton or to the neutron state for which the
standard interpolating fields are given by the formulae:
\be
\mcJ_p = \epsilon^{abc} \left( u^{a,T} \mcC\gamma_5 d^b \right) u^c \gap\textmd{and}\gap   \mcJ_n = \epsilon^{abc} \left( d^{a,T} \mcC\gamma_5 u^b \right) d^c.
\ee
Note that using discrete symmetries and anti-periodic boundary conditions in the
time direction for the quark fields, one can show that
$C^{+}_{N,\rm 2pts}(t)  = -C^{-}_{N,\rm 2pts}(T-t)$.
Let us also recall that an exact symmetry of the action leads to the
following relation at a finite value of lattice spacing,
$C^{\pm}_{n,\rm 2pts} (t) =  C^{\pm}_{p,\rm 2pts} ( t )$ \cite{Alexandrou:2008tn}.
In order to improve the overlap between the ground state and the
interpolating fields of the nucleon we use Gaussian smearing of the quark fields
appearing in the interpolating fields. We also use APE smearing of the
gauge links involved in the Gaussian smearing.

The nucleon three-point functions is given by
\be\label{eq:C3pts}
C^{\pm,a}_{N,\rm 3pts}(t_{s},t_{\rm{op}}) =     \sum_{\vec{x_{\rm s}} ,\vec{x}_{\rm op}}   \rm{tr}~ \Gamma^{\pm}\la \mcJ_{N}(x_{\rm s}) O^{a=3,8}_{\{44\}}(x_{\rm op}) \overline{\mcJ_{N}}(0)  \ra ,
\ee
where  $ O^{a=3,8}_{\{44\}}$ is one of the twist-2 operators introduced
in \eq{eq:twist2_unpol}, $t_{\rm op}$ is the time of insertion of the
operator, and $t_s$ denotes the so-called source-sink separation.
Note that the precise definition of the strange quark field entering in the
operator  $O^{a=8}_{\{44\}}$ is postponed to the next subsection.

Using the two- and three-point correlators of~\eqs{eq:C2pts}
and~(\ref{eq:C3pts}), we construct the following ratio:

\be\label{eq:ratio_def}
R^{a}(t_s,t_{\rm op}) \equiv \frac{C^{+,a}_{N,\rm 3pts}(t_s,t_{\rm op})}{ am_N C^{+}_{N,\rm 2pts}(t_s) }  =  \la x \ra^{(a)}_{\rm bare} + \mcO( e^{-\Delta M  t_{\rm op}}) +  \mcO( e^{-\Delta M (t_s - t_{\rm op})} ),
\ee
where $am_N$ is the nucleon mass in lattice units and $\Delta M$ is the mass
gap between the lowest nucleon state and the
first excited state with the same quantum numbers.
One can thus extract from the asymptotic time behaviour of  $R^{a=8}(t_s,t_{\rm op})$
the bare quantity $\la x \ra^{(8)}_{\rm bare}$ and correspondingly $\la x \ra^{(3)}_{\rm bare}$.

\subsection{Lattice evaluation}
\label{subsec:lat_eval}

While the light quark fields used in the operator $O^{a=3,8}_{\{44\}}$ are the
unitary fields we use, as mentioned already above, a different action
for the valence strange quark. In practice we introduce a doublet of
mass degenerate quarks with a mass $a\mu_s$ tuned to reproduce the unitary Kaon mass.
This procedure introduces an error due to the uncertainty on the determination of
the matching mass that we will discuss later on but will allow us to use an
efficient noise reduction technique that will be explained in the next section.

\begin{figure}
\centering
\mbox{\subfigure{\includegraphics[width=0.425\linewidth]{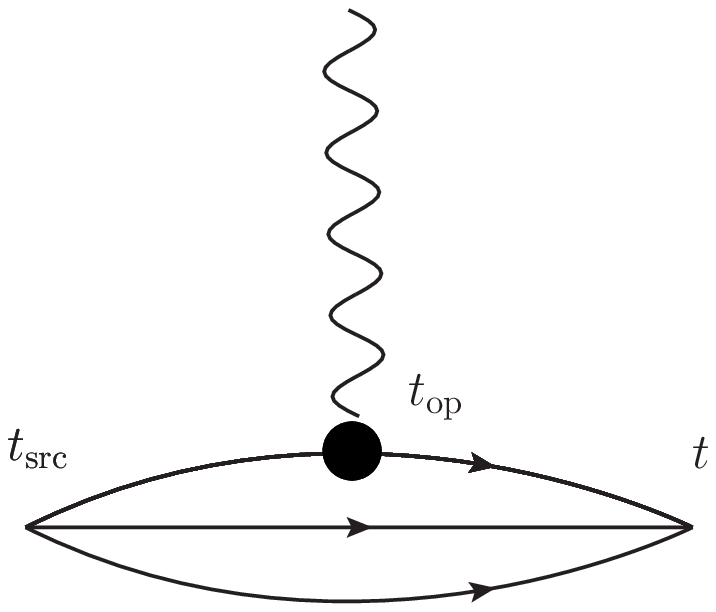}}\label{fig:contract_conn}
\quad
\subfigure{\includegraphics[width=0.425\linewidth]{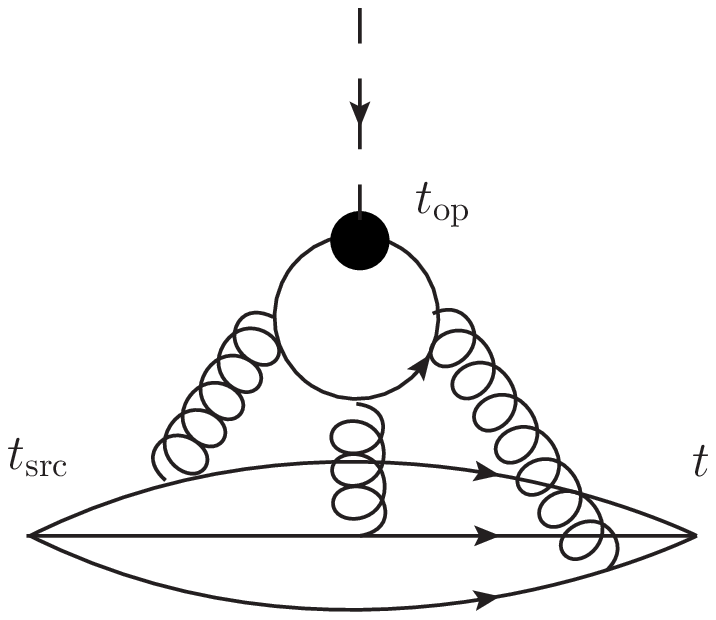}}\label{fig:contract_disc} }
\caption{Connected (left) and the disconnected
  (right) graphs arising from the Wick contractions of the here considered 3-point functions.}
\label{fig:contract}
\end{figure}

Consider the following operator in terms of the field in the twisted mass basis :
\be
J^{8} = \chibar  \gamma_{\{4} i \overleftrightarrow{D}_{4\}}   \chi - \bar{\chi}_s \gamma_{\{4} i \overleftrightarrow{D}_{4\}}  \chi_s
\ee
Performing the rotation to the physical basis, we obtain :
\be
J^{8} = \psibar  \gamma_{\{4} i \overleftrightarrow{D}_{4\}}   \psi - \bar{\psi}_s \gamma_{\{4} i \overleftrightarrow{D}_{4\}}  \psi_s = \bar{u} \gamma_{\{4} i \overleftrightarrow{D}_{4\}} u +  \bar{d} \gamma_{\{4} i \overleftrightarrow{D}_{4\}} d - \bar{s}_+ \gamma_{\{4} i \overleftrightarrow{D}_{4\}} s_+ - \bar{s}_- \gamma_{\{4} i \overleftrightarrow{D}_{4\}} s_-
\ee
Note that $J^{8}$ keeps the same form in the two bases and that $J^{8}$
is only one possible choice for a discretization of the operator  $O^{a=8}_{\{44\}}$.
While the two-point nucleon correlators of \eq{eq:C2pts} give only rise to
quark-connected Wick contractions, in general the three-point functions of
\eq{eq:C3pts} yield both quark-connected (illustrated in Fig.~\ref{fig:contract}a)
and quark-disconnected (illustrated in Fig.~\ref{fig:contract}b) contributions.
In the following we will refer to them simply as to connected and disconnected
 fermionic Wick contractions (or diagrams) and shall write
\be\label{eq:3pts}
C^{\pm,a}_{N,\rm 3pts}(t_{s}, t_{\rm{op}}) =  \widetilde{C}^{\pm,a}_{N,\rm 3pts}(t_{s}, t_{\rm{op}}) +  \mcD^{\pm,a}_{N,\rm 3pt}(t_s,t_{\rm{op}})
\ee
with $\widetilde{C}^{\pm,a}_{N,\rm 3pt}$ (resp.\ $\mcD^{\pm,a}_{N,\rm 3pt}$)
corresponding to the connected (resp.\ disconnected) quark diagrams, defined as
\ba
\hspace{-1.cm}&&\widetilde{C}^{\pm,a}_{N,\rm 3pt}(t_s,t_{\rm{op}}) = \!
\sum_{\vec{x} ,\, \vec{x}_{\rm op}} {\rm{tr}}~\left\{\Gamma^{\pm}\la  \big[ \mcJ_{N}(x) O^{a}_{\{44\}}(x_{\rm op})  \overline{\mcJ_{N}}(x_{\rm src})  \big]
\ra\right\} \, ,
\label{eq:C3pts-Wick} \\
\hspace{-1.cm}&&\mcD^{\pm,a}_{N,\rm 3pt}(t_s,t_{\rm{op}}) =\!
\sum_{\vec{x},\,\vec{x}_{\rm op}}  {\rm{tr}}~\left\{
\Gamma^{\pm}  \la
\big[ \mcJ_{N}(x) \overline{\mcJ_{N}}(x_{\rm src}) \big]
\big[  O^{a}_{\{44\}}(x_{\rm op}) \big]
\ra\right\} \, ,
\label{eq:D3pts-Wick}
\ea
where the symbol $[...]$ is a shorthand for all the {\em connected}
fermionic Wick contractions.
Note that for $a=3$, the disconnected part is a $\mcO(a^2)$ effect which vanishes
in the continuum limit and can thus be neglected.
Introducing
\be
\delta_{\pm}^{(\mu,\mu_s)}(t_{\rm op}) =  \sum_{\vec{x}_{\rm op}} {\rm{tr}}~ ~\left\{ \left ( \frac{1}{D_{l,\pm}[U]}  -
 \frac{1}{D_{s,\pm}[U]}  \right )_{(x_{\rm op},x_{\rm op})}~\right\}   \, ,  \label{eq:delta_pm}
\ee
the contribution of the disconnected fermion loop
to $\mcD^{\pm,a}_{N,\rm 3pt}$ on
a given gauge configuration $U$
in our setup reads
\be\label{eq:disc_part}
  \mcD^{\pm,a}_{N,\rm 3pt}(t_s,t_{\rm{op}}) = \la C^{\pm }_{N,\rm 2pts}(t_s)  \left(\delta_{+}^{(\mu,\mu_s)}(t_{\rm op}) +\delta_{-}^{(\mu,\mu_s)}(t_{\rm op})\right)  \ra,
\ee

The connected contributions $\widetilde{C}^{\pm,a}_{N,\rm 3pt}(t_s,t_{\rm{op}})$
have been evaluated using standard techniques for three-point functions
(sequential inversions through the sink), see e.g. ref.~\cite{Alexandrou:2013joa}.

\subsection{Estimation of disconnected loops}
\label{subsec:hl_method}

We describe here the generalization of the variance reduction method for twisted mass fermions
introduced and discussed in ~\cite{Michael:2007vn,Boucaud:2008xu,Dinter:2012tt,Alexandrou:2013wca}.

Consider the identity

\be
D_{l,\pm} - D_{s,\pm} = \pm  i \gamma_5  a \left( \mu_l  - \mu_{s}\right)
\label{eq:diff_Dq_Dqp}
\ee
implying that
\be\label{eq:vv_hl_method}
\frac{1}{D_{l,\pm}} - \frac{1}{D_{s,\pm}} = -\frac{1}{D_{s,\pm}} \left(D_{l,\pm} - D_{s,\pm}\right)  \frac{1}{D_{l,\pm}} =   \mp  i    a \left( \mu_l  - \mu_{s}\right) \frac{1}{D_{s,\pm}}  \gamma_5   \frac{1}{D_{l,\pm}},
\ee
where we have used \eq{eq:diff_Dq_Dqp} to obtain the last equality.

For the practical calculation, we introduce a set $\Xi$ of $N_\xi$ independent
random volume sources, $\{\xi_{[1]},\dots, \xi_{[r]}, \dots, \xi_{[N_\xi]}\}$, satisfying
\be
\lim_{N_\xi \to \infty}\left[ \xi^{  i}_{ [r]}(x)^\ast \xi^j_{ [r]}(y) \right]_\Xi
= \delta_{xy} \delta^{ij}
\label{eq:Rsources}
\ee
where $i=1,...,12$ refers to the spin and color indices of the
source and $[ \dots]_\Xi$ denotes the average over the $N_\xi$
noise sources in $\Xi$ .

Applying $O^{a}_{\{44\}}$ to \eq{eq:vv_hl_method} and
taking the trace over spin and colour indices we obtain :
\be\label{eq:vv_hl_method_dis}
\mp i a \left( \mu_l  - \mu_{s}\right) \sum_{\vec{x}} \left[ \phi_{[r],l,\mp}^* \gamma_5  O^{a}_{\{44\}} \phi_{[r],s,\pm} (x) \right]_R  = \delta_{\pm}^{(\mu,\mu_s)}(t_{\rm op}) + \mcO\left( R^{-1/2}\right),
\ee
where
\be
 \phi_{[r],s,\pm}=(1/D_{s,\pm}) \xi_{[r]} \gap\textmd{and}\gap \phi_{[r],l,\pm}^* = \xi_{[r]}^* (1/D_{l,\pm})^{\dag}.
\ee

For the generation of the random sources we have used
a $\mathbf{Z}_2$ noise taking
all field components randomly from the set $\{1,-1\}$.

Note that $\delta_{\pm}^{(\mu,\mu_s)}$, and by construction its variance,
is proportional to the mass difference $\mu_l - \mu_s$ and vanishes on
each configuration in the limit $ \mu_s \to \mu_l$.  Our approach, thus,
exactly encodes the fact that the disconnected contributions we are interested in
vanish in the $SU(3)_{\rm flavour}$ limit.

\subsection{Renormalization}

The renormalization of the  operator  $O^{a=3}$ is  known to be multiplicative
from our previous work \cite{Alexandrou:2013joa}  and have been obtained
non perturbatively using the methodology developed in
\cite{Alexandrou:2010me}. 
The renormalization factor  $Z^{\mu\mu}_{DV}(\beta)$ read :
\be\label{eq:Z44}
Z^{\mu\mu}_{DV}(\beta =1.95) = 1.019(4) ,\gap Z^{\mu\mu}_{DV}(\beta =2.10) = 1.048(5)\; .
\ee
Note that an independent calculation
performed in \cite{Blossier:2014kta,Blossier:2014rga} gives compatible results.
In the limit $\mu_s=\mu_l=0$, $SU(3)_{\rm flavour}$ is an exact symmetry of the action
and the operators $O^{a=3,8}_{\{\mu\nu\}}$ belong to the same flavour multiplet.
They thus share the same renormalization pattern in a mass independent scheme.
The ratio  $\frac{\la x \ra^{(3)}}{\la x \ra^{(8)}}$ is thus renormalization free.

\section{Results}

\begin{figure}[htb]
\begin{center}
\includegraphics[width=300pt,angle=\plotangle]{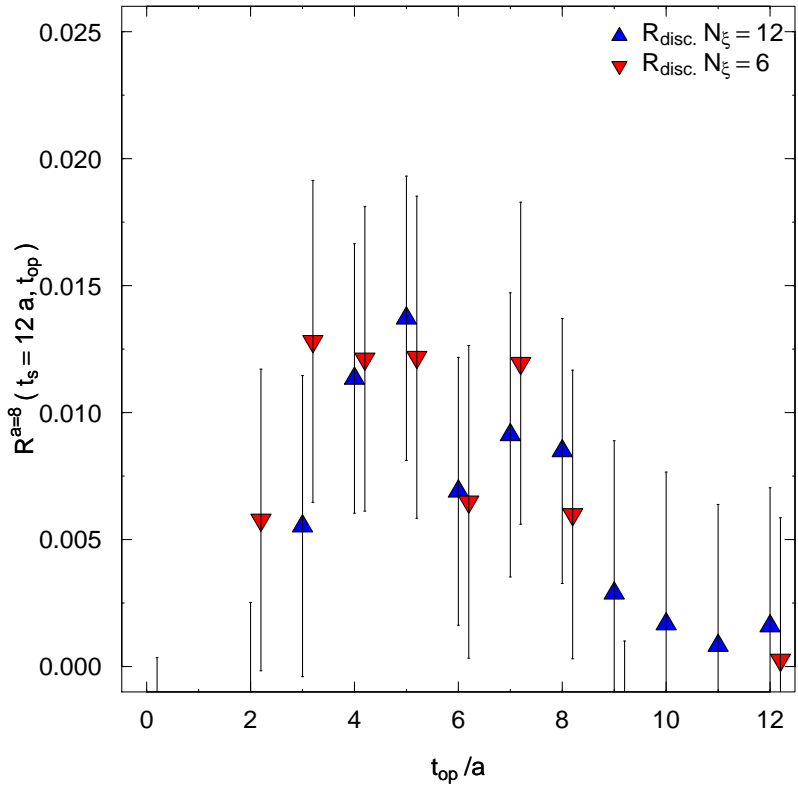}
\end{center}
\caption{Plot of the contributions $R_{\rm{disc.}}$ versus  $t_{\rm op}$ at $t_s=12a$
for $N_\xi=6$ (red down triangle) and $N_{\xi}=12$ (blue up triangle).
The bare mass of the strange quark is fixed to $a\mu_s=0.018$ and we use
$a\mu_l=0.0055$ and $\beta=1.95$  }\label{fig:disc_plateau_x8_ts12}
\end{figure}

As a first step, we have investigated the magnitude of the stochastic noise
introduced by the method described in section~\ref{subsec:hl_method}. To this end, we used a
fixed number of gauge configurations for a gauge field ensemble at
coupling $\beta=1.95$ and twisted mass parameter $a\mu_l=0.0055$.
We show in \fig{fig:disc_plateau_x8_ts12}, the ratio  $R^{a=8}_{\rm{disc.}}$ for a
fixed source-sink separation of $t_s\sim 1~\fm$ as a function of $t_{\rm op}$
for $N_\xi=6$ and $N_\xi=12$. As can be seen, the signal is compatible with zero
within the errors when using
$N_\xi=12$. Furthermore we observe that the error does not decrease much
when the number of stochastic sources is doubled. This means presumably that the
error is already dominated by the intrinsic noise of the gauge field fluctuations.
We nevertheless decided to use  $N_\xi=12$ throughout this
benchmark study of the octet moment.

\begin{figure}[htb]
\begin{center}
\includegraphics[width=300pt,angle=\plotangle]{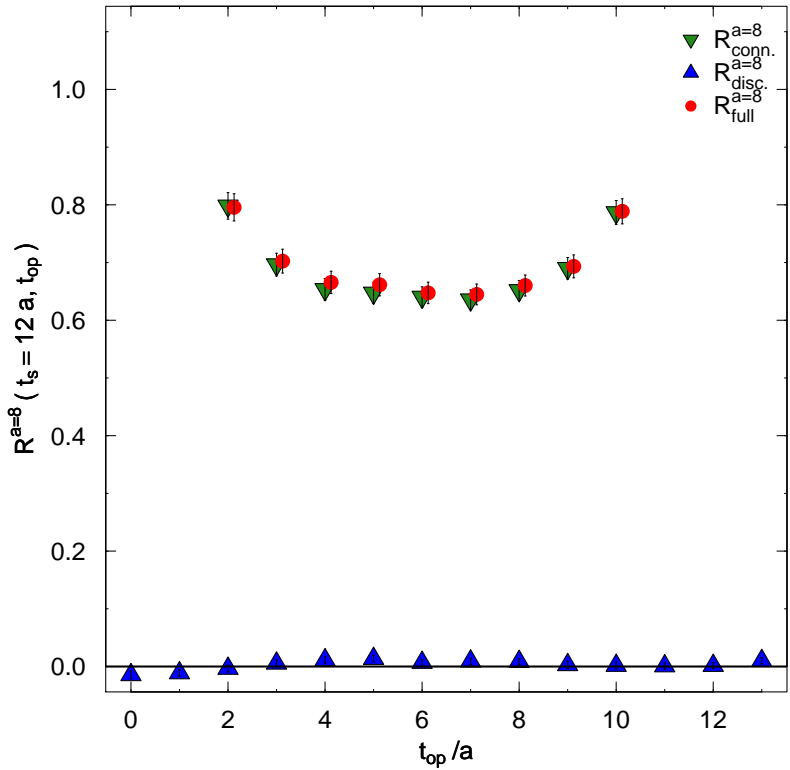}
\end{center}
\caption{Plot of the contributions $R^{(8)}_{\rm{disconnected}}$
(blue triangle), $R^{(8)}_{\rm{connected}}$
(black triangles) and of their sum, $R^{(8)}_{\rm full}$ (red filled circles) as
function of $t_{\rm op}$ at $t_s=12a$ for  $a\mu_l=0.0055$ and $\beta=1.95$.}
\label{fig:plateau_x8_ts12}
\end{figure}

In principle, it would therefore be possible to reduce the error
further by using more stochastic noise vectors. However,
the contribution of the disconnected 3-point function is
small compared to the value of the connected part as demonstrated in
\fig{fig:plateau_x8_ts12} where we show the connected contribution
$R^{a=8}_{\rm{connected}}(t_s=12a,t_{\rm op}) $, the disconnected
contribution  $R^{a=8}_{\rm{disconnected}}(t_s=12a,t_{\rm op})$
and the full correlator $R^{a=8}_{\rm{full}}(t_s=12a,t_{\rm op})$.
In particular, the statistical errors on the connected part are about $~0.016$ ($\sim 2.5\%$) which is significantly larger than the size of
the disconnected contributions.
Despite this fact, which would make neglecting the dis-connected contributions
tempting,
we always include them in the following analysis.
Note that the $R^{a=8}_{\rm{disconnected}}$ is proportional to the
difference between the light and strange quark mass. Since we are
using a mixed action setup, our results depends on an approximate
tuning of the strange quark mass. In all the figures we use $a\mu_s(\beta=1.95)=0.018$
for $\beta=1.95$ and $a\mu_s(\beta=2.10)=0.015$ for $\beta=2.10$.f
Those values can be
compared to the values obtained for instance in \cite{Alexandrou:2014sha} where the
strange quark mass has been determined using the $\Omega^-$ mass and
correspond to $a\mu_s(\beta=1.95) = 0.0194 $ and
$a\mu_s(\beta=2.10)=0.0154$.
However, by using data
on the $\beta=1.95$ ensemble we have explicitly checked that changing the bare strange quark mass
by more than $\sim 10\%$ does not lead to any significant change in the value of the
disconnected contribution.

\subsection{Excited states contamination}
\label{subsec:excited}

\begin{figure}[h]
\begin{center}
\includegraphics[width=300pt,angle=\plotangle]{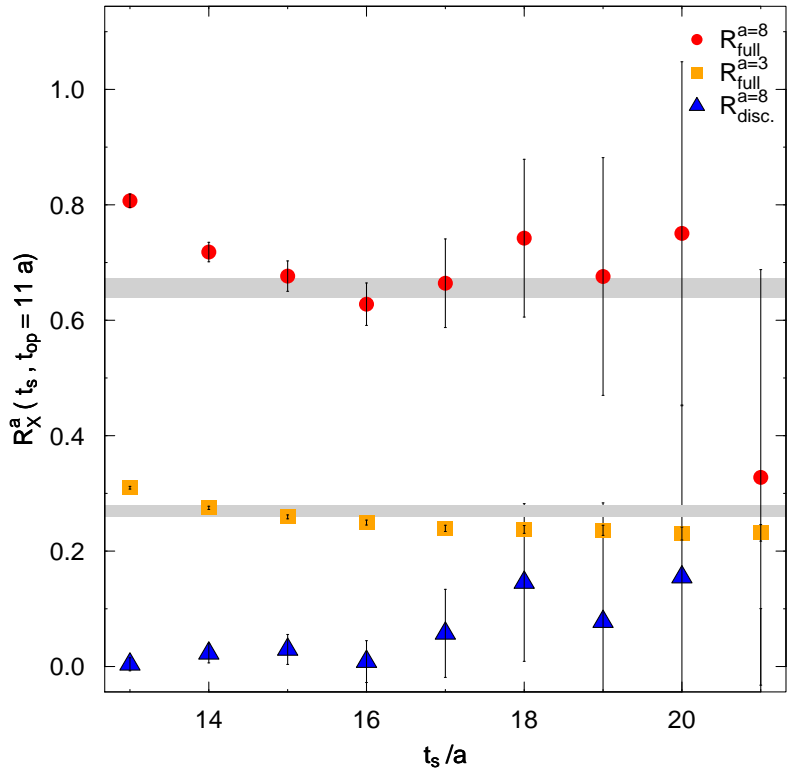}
\end{center}
\caption{ The ratio \eq{eq:ratio_def} as a function of the source-sink
separation $t_s$ for a fixed source to operator time $t_{\rm op} = 11a$ in the
isovector (orange squares) and octet case (red dots).
The gray bands indicate the results obtained from a fixed sink
calculation for $t_s=12a$.
The blue triangles show the disconnected contribution to $R^{a=8}(t_s,t_{\rm op}=11a)$. }\label{fig:opensink_x8}
\end{figure}

In order to investigate the contamination of excited states due to the second
and third term in \eq{eq:ratio_def}, we used the same procedure ("open-sink'' method)
as in \cite{Dinter:2011sg}, namely we study the source-sink dependence for a
fixed source to operator  separation $t_{\rm op}\sim 0.9~\fm$ ( $t_{\rm op}=11a$).
Details on the technical implementation of the "open-sink'' method can be
found in \cite{Dinter:2011sg}.  To this end, a large statistics
for the connected part ($\sim 23000$ measurements) has been used.
We plot in \fig{fig:opensink_x8}
the resulting renormalized ratios $R^{a=3,8}(t_s,t_{\rm op}=11a)$.
The results in the isovector case are represented by  orange squares and the
results in the flavour octet case (including the disconnected piece) are
depicted by red dots. We also represent using blue triangles the disconnected
contribution to $R^{a=8}(t_s,t_{\rm op}=11a)$. As can be seen, the noise
for the dis-connected part
dominates for large source-sink separation. Nevertheless, we can obtain
a reasonable signal up to source-sink separation of about $16a$
($\approx 1.3 ~\fm$). The gray bands indicate the values  of $\la x^{(a=3,8)} \ra $
obtained using a fixed source-sink separation calculation with $t_s
\sim 1~\fm$. Note that the fitting ranges have been determined choosing
the fit with the longest plateau with a confidence level of least $90\%$.
In the flavour octet case we observe that results obtained for $t_s/a
> 15$ and a fixed source-operator separation $t_{\rm op} = 11a$ are compatible with the result obtained at a fixed source-sink
separation of $t_s/a=12$. We mention that at the here used value of the pion mass of $m_{\rm{PS}}\approx 370 ~\mev$, volume $(L=2.6~\fm)$
and lattice spacing $(a\approx 0.082~\fm)$ we found  for the isovector
channel a relative shift of about $10\%$ for a corresponding
comparison at different source sink separations\cite{Dinter:2011sg}.

\begin{figure}[h]
\begin{center}
\includegraphics[width=300pt,angle=\plotangle]{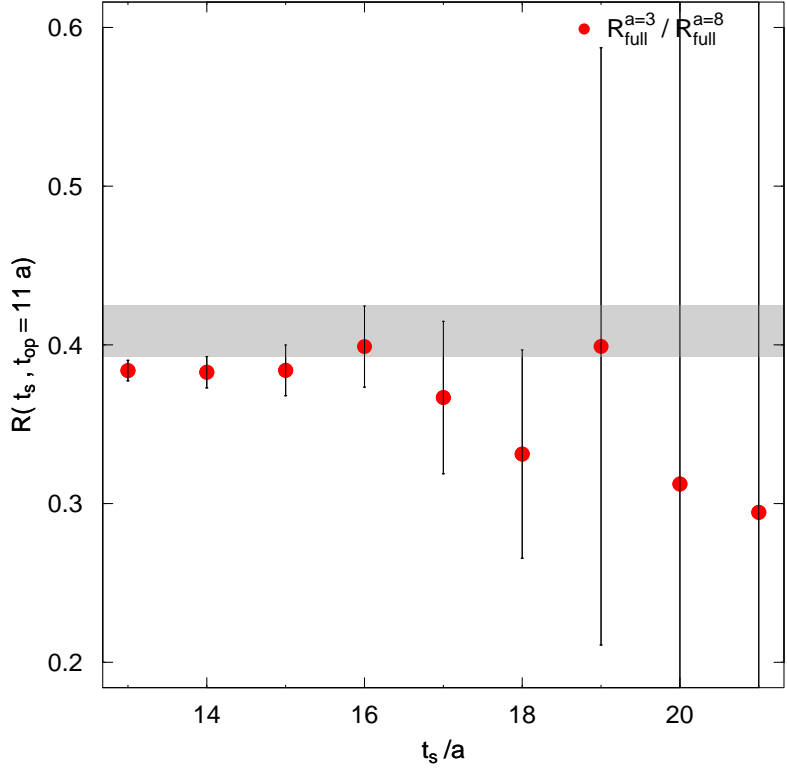}
\end{center}
\caption{ $\frac{R^{a=3}(t_s,t_{\rm op}=11a)}{R^{a=8}(t_s,t_{\rm op}=11a)}$ as a
function of the source-sink separation $t_s$ for a fixed source to operator time
$t_{\rm op} = 11a$ (red dots). The gray band indicates the result
obtained from a fixed sink calculation of $t_s=12a$.  }\label{fig:opensink_ratio_x3_x8}
\end{figure}

We performed a similar study of the ratio of the octet to the singlet
3-point functions of~\eq{eq:3pts},
\be
R(t_s,t_{\rm op})=C^{a=3}_{N,\rm 3pts}(t_s,t_{\rm op})/C^{a=8}_{N,\rm
  3pts}(t_s,t_{\rm op}).
\ee
The ratio $R(t_s,t_{op}=11a)$ is shown in
\fig{fig:opensink_ratio_x3_x8} as a function of the source-sink separation
$t_s$.The constant fit for this ratio obtained with a fixed source to
operator time
$t_{\rm op}=11a$ is depicted by a gray band. We used the same criterion as in
\fig{fig:opensink_x8} to determine the fitting range.
In this case we do not observe any significant source-sink dependence
of our results and thus there seem to be only a small excited
states contamination, at least within the size of our errors. Note  that the statistical errors stemming from
the disconnected contribution are responsible for most of the total
statistical error at large source-sink separation. In the next section
we will use the difference between the central value at $t_s \sim
1.3\fm$ with the open think method and the central value obtained with
the fixed source-sink method at $t_s \sim 1\fm$ as an estimate of the
systematical error for $\la x^{(a=3,8)} \ra$ and $\frac{\la x^{(3)} \ra }{\la x^{(8)} \ra } $.

\newpage

\subsection{Chiral behaviour}

\begin{figure}[h]
\begin{center}
\includegraphics[width=300pt,angle=\plotangle]{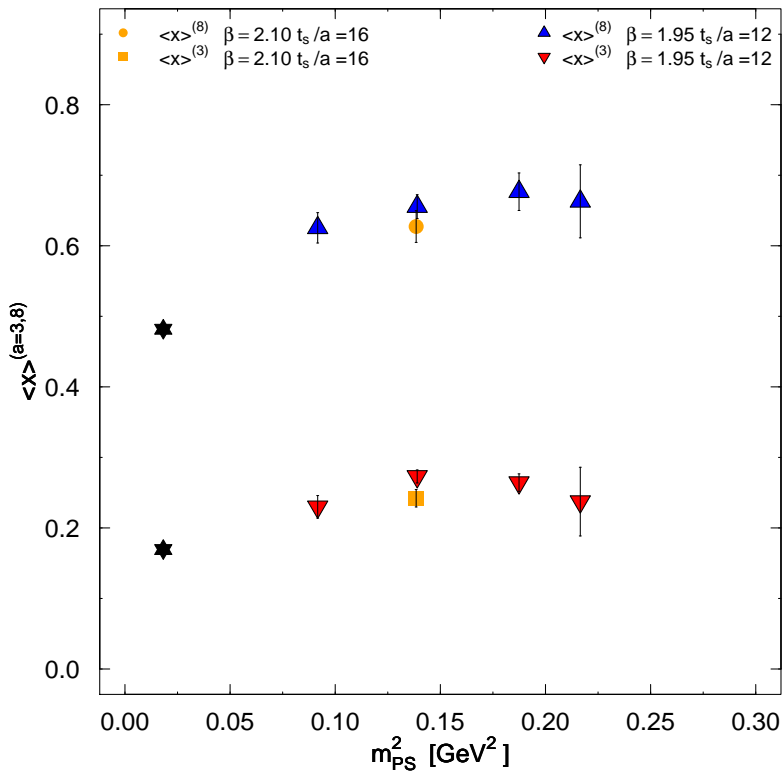}
\end{center}
\caption{$\la x^{(3,8)} \ra_{\mu^2 = 4 ~\gev^2} $ as a function of the pseudoscalar
meson mass $\mps^2$. The phenomenological estimates are represented by two black stars. }
\label{fig:xfit_x8}
\end{figure}

We plot our results for  $\la x^{(3,8)} \ra_{\mu^2 = 4 ~\gev^2} $  as a function
of the pseudoscalar meson mass $\mps^2$ in \fig{fig:xfit_x8}.
The results for $\la x^{(3)} \ra_{\mu^2 = 4 ~\gev^2} $ (respectively
$\la x^{(8)} \ra_{\mu^2 = 4 ~\gev^2} $) are shown using red down
triangles (resp. blue up triangle) for a lattice spacing of $0.082~\fm$
and by an orange square (resp. an orange dot) for a lattice spacing of $0.064~\fm$.
Neglecting, in a first step, excited state contaminations and
in order to investigate the chiral limit behaviour
of our data, all the results have been obtained using a fixed source-sink separation of
approximately $1~\fm$.
Of course, in our final result we will add the shift
from the excited states contamination as a systematic error.
As can be seen in  \fig{fig:xfit_x8}
the results obtained at two different lattice spacings are
compatible, and $\mcO(a^2)$ effects can be neglected.

The phenomenological estimates \cite{Sergey} of \eq{eq:pheno_x} are represented by two
black stars.
As can be seen in the graph, for both, $\la x^{(3)} \ra$ and $\la x^{(8)} \ra$
and for unphysically large values of the pion mass
the lattice data lay consistently above the phenomenological values,
a phenomenon that is well know from previous investigations
of $\la x^{(3)} \ra$, see e.g. \cite{Constantinou:2014tga},
and is here demonstrated for the first time for $\la x^{(8)} \ra$.

As explained before, systematic errors stemming from discretization effects
and from the matching mass dependence can be neglected. With the present data
set we cannot estimate safely the volume dependence of $\la x^{(3,8)} \ra$,
however knowing that they are negligible in the isovector case in our $N_f=2$
calculation \cite{Alexandrou:2011nr} we assume them to be small and neglect them, too.
Our dominating source of systematic error for the
individual matrix elements is then due to the excited states contamination.
In addition, as the chiral extrapolation is not performed here,
it remains as a presently unquantifiable systematic error.

\begin{figure}[h]
\begin{center}
\includegraphics[width=300pt,angle=\plotangle]{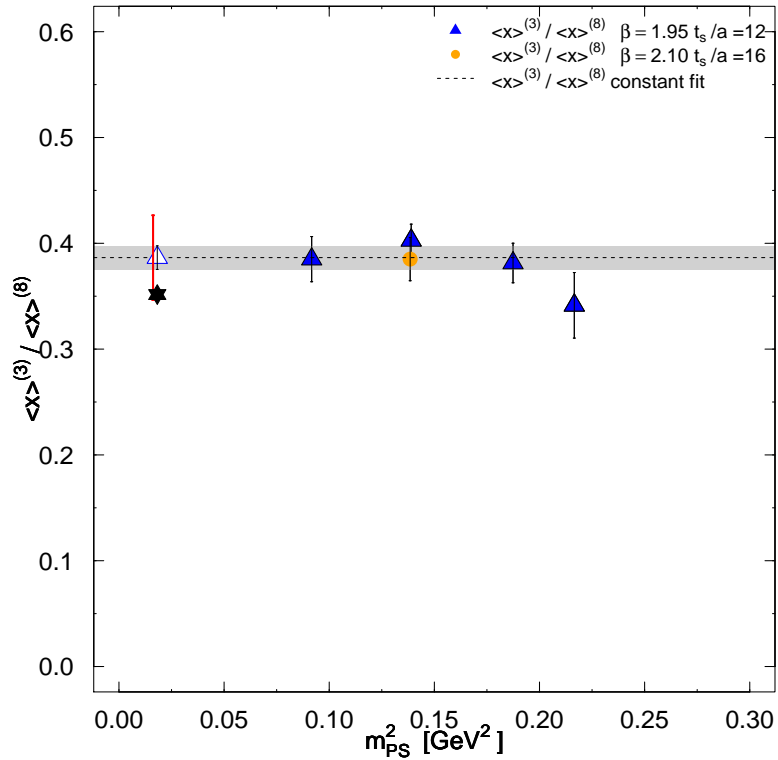}
\end{center}
\caption{$\la x^{(a=3)} \ra_{\mu^2 = 4 ~\gev^2}/\la x^{(a=8)} \ra_{\mu^2 = 4 ~\gev^2} $
as a function of the pseudoscalar meson mass $\mps^2$ for two values
of the lattice spacing. The result of a constant extrapolation is
represented by an empty triangle. The systematic error on the
extrapolated value is represented by a red error bar slightly shifted
for readability. The phenomenological estimate from is represented by a black star.}\label{fig:xfit_x3_over_x8}
\end{figure}

In fig.~\ref{fig:xfit_x3_over_x8} we show the ratio $\frac{\la x^{(3)} \ra}{ \la x^{(8)} \ra}$
as a function of the pion mass together with the phenomenological value.
Here the situation is somewhat different in that the lattice data agree with
the phenomenological analysis even at the here used unphysically large pion
masses. This suggests that some of the systematic uncertainties cancel
in the ratio.
As discussed in section~\ref{subsec:excited},
the systematic error stemming from the excited states
contamination is negligible for the ratio $\frac{\la x^{(3)} \ra}{ \la x^{(8)} \ra}$.
Therefore, we conclude for the individual moments that
either the non-perturbative renormalization or the chiral extrapolation
is the most probable systematic effect leading to the observed discrepancies
in $ \la x^{(3,8)} \ra$. It would therefore be highly desirable to
perform simulations directly at the physical point to eliminate the uncertainty from
the chiral extrapolation.
Work in this direction is in progress \cite{Abdel-Rehim:2013yaa,Abdel-Rehim:2014nka}.

Since the lattice data are flat as function of the pion mass, we perform a simple
constant extrapolation to the physical pion mass. As a systematic error, we take the difference between
the data points at the smallest and the largest pion mass. We find
then for $\la x^{(a=3)} \ra_{\mu^2 = 4 ~\gev^2}/\la x^{(a=8)} \ra_{\mu^2 = 4 ~\gev^2}$

\be
\frac{\la x^{(3)} \ra}{ \la x^{(8)} \ra}  = 0.39(1)(4)\;
\label{eq:ratios}
\ee
where the first error is statistical and the second one is our
estimate of the systematic error. When comparing to phenomenological extractions \cite{Sergey}
we find that the ratio $\frac{\la x^{(3)} \ra}{ \la x^{(8)}}$
is, within errors, compatible with phenomenological extractions at least
for the here used simple extrapolation to the physical point.

\section{Conclusion}

In this work we have performed a benchmark computation of the flavour
octet combination of the first moment of parton distribution functions
of the nucleon using $N_f=2+1+1$ twisted mass fermions tuned to
maximal twist. This quantity can provide a first hint of the contribution of
the strange quark to the quark momentum in the nucleon.
We have utilized the fact that $\xoct_{\mu^2}$ shares the same renormalization
property as the standard iso-vector moment $\xiso_{\mu^2}$. Using our
result for the non-perturbative renormalization constant obtained in \cite{Alexandrou:2013joa} we can
thus provide an $\mcO(a^2)$ improved estimate of $\xoct_{\mu^2}$. The
calculation takes into account the notably noisy dis-connected
contribution, which are shown to have a negligible contribution in our
setup.

In our investigation of $\xoct_{\mu^2}$ we found, similar to the case of
$\xiso_{\mu^2}$ \cite{Dinter:2011sg}, that
the excited states contamination
can be substantial and has to be taken into account as a systematic error.
We also found that, similar to the case of $\xiso_{\mu^2}$,
the lattice data for $\xoct_{\mu^2}$ are systematically
higher than the phenomenological value, see fig.~\ref{fig:xfit_x8} for
pion masses larger than the physical one.

However, when the ratio
$\la x^{(a=3)} \ra_{\mu^2 = 4 ~\gev^2}/\la x^{(a=8)} \ra_{\mu^2 = 4 ~\gev^2}$
is considered, the lattice data show an overall agreement with the corresponding
phenomenological ratio, indicating that in the ratio
systematic errors cancel. Since in the ratio the excited states contamination
is small, the here observed agreement cannot be due to
this systematic uncertainty. One reason why the ratio agrees with the
phenomenological value can be
that the renormalization constants cancel. Another possibility is that both,
$\xoct_{\mu^2}$ and $\xiso_{\mu^2}$ have a very similar chiral
limit behaviour such that the effects from the chiral extrapolation cancel.
It would be highly desirable to have therefore lattice results directly at the
physical point for both quantities.
Since the lattice data are rather flat, we performed a simple constant extrapolation
to the physical point for the ratio
$\la x^{(a=3)} \ra_{\mu^2 = 4 ~\gev^2}/\la x^{(a=8)} \ra_{\mu^2 = 4 ~\gev^2}$
and find
\be
\frac{\la x^{(3)} \ra}{ \la x^{(8)} \ra}  = 0.39(1)(4)
\ee
This value is in agreement with the phenomenological extraction of this
quantity from deep inelastic scattering data \cite{Sergey}.

\section*{Acknowledgments}

We are particularly
grateful to S. Alekhin for having provided us with the
phenomenological estimates of $\xoct$. We thank our fellow members of
ETMC for their constant collaboration and S. Alekhin and J. Bl\"umlein for useful discussions. We are grateful to the John von Neumann Institute for Computing (NIC),
the J{\"u}lich Supercomputing Center and the DESY Zeuthen Computing
Center for their computing resources and support.  This work has been
supported in part by the DFG Sonderforschungsbereich/Transregio
SFB/TR9.  This work was supported by the Danish National Research
Foundation DNRF:90 grant and by a Lundbeck Foundation Fellowship
grant.This work is supported in part by the Cyprus Research Promotion Foundation under contracts KY-$\Gamma$/0310/02/, and the Research Executive Agency of the European Union under Grant Agreement number PITN-GA-2009-238353 (ITN STRONGnet).
K. J. was supported in part by the Cyprus Research Promotion
Foundation under contract $\Pi$PO$\Sigma$E$\Lambda$KY$\Sigma$H/EM$\Pi$EIPO$\Sigma$/0311/16.

\bibliography{octet_current}

\providecommand{\href}[2]{#2}\begingroup\raggedright\begin{thebibliography}{10}

\bibitem{Dinter:2012tt}
{\bf ETM Collaboration} Collaboration, S.~Dinter {\em et.~al.}, {\it {Sigma
  terms and strangeness content of the nucleon with $N_f=2+1+1$ twisted mass
  fermions}},  {\em JHEP} {\bf 1208} (2012) 037,
  [\href{http://xxx.lanl.gov/abs/1202.1480}{{\tt arXiv:1202.1480}}].

\bibitem{Alexandrou:2013joa}
C.~Alexandrou, M.~Constantinou, S.~Dinter, V.~Drach, K.~Jansen, {\em et.~al.},
  {\it {Nucleon form factors and moments of generalized parton distributions
  using $N_f=2+1+1$ twisted mass fermions}},  {\em Phys.Rev.} {\bf D88} (2013)
  014509, [\href{http://xxx.lanl.gov/abs/1303.5979}{{\tt arXiv:1303.5979}}].

\bibitem{Alexandrou:2011nr}
C.~Alexandrou, J.~Carbonell, M.~Constantinou, P.~Harraud, P.~Guichon, {\em
  et.~al.}, {\it {Moments of nucleon generalized parton distributions from
  lattice QCD}},  {\em Phys.Rev.} {\bf D83} (2011) 114513,
  [\href{http://xxx.lanl.gov/abs/1104.1600}{{\tt arXiv:1104.1600}}].

\bibitem{Alexandrou:2011db}
C.~Alexandrou, M.~Brinet, J.~Carbonell, M.~Constantinou, P.~Harraud, {\em
  et.~al.}, {\it {Nucleon electromagnetic form factors in twisted mass lattice
  QCD}},  {\em Phys.Rev.} {\bf D83} (2011) 094502,
  [\href{http://xxx.lanl.gov/abs/1102.2208}{{\tt arXiv:1102.2208}}].

\bibitem{Alexandrou:2010hf}
{\bf ETM Collaboration} Collaboration, C.~Alexandrou {\em et.~al.}, {\it {Axial
  Nucleon form factors from lattice QCD}},  {\em Phys.Rev.} {\bf D83} (2011)
  045010, [\href{http://xxx.lanl.gov/abs/1012.0857}{{\tt arXiv:1012.0857}}].

\bibitem{Bali:2013dpa}
G.~Bali, S.~Collins, B.~Gläßle, M.~Göckeler, J.~Najjar, {\em et.~al.}, {\it
  {Nucleon generalized form factors and sigma term from lattice QCD near the
  physical quark mass}},  \href{http://xxx.lanl.gov/abs/1312.0828}{{\tt
  arXiv:1312.0828}}.

\bibitem{Bali:2012av}
G.~S. Bali, S.~Collins, M.~Deka, B.~Glassle, M.~Gockeler, {\em et.~al.}, {\it
  {$\langle x \rangle_{u-d}$ from lattice QCD at nearly physical quark
  masses}},  {\em Phys.Rev.} {\bf D86} (2012) 054504,
  [\href{http://xxx.lanl.gov/abs/1207.1110}{{\tt arXiv:1207.1110}}].

\bibitem{Green:2012ud}
J.~Green, M.~Engelhardt, S.~Krieg, J.~Negele, A.~Pochinsky, {\em et.~al.}, {\it
  {Nucleon Structure from Lattice QCD Using a Nearly Physical Pion Mass}},
  \href{http://xxx.lanl.gov/abs/1209.1687}{{\tt arXiv:1209.1687}}.

\bibitem{Hagler:2007xi}
{\bf LHPC Collaborations} Collaboration, P.~Hagler {\em et.~al.}, {\it {Nucleon
  Generalized Parton Distributions from Full Lattice QCD}},  {\em Phys.Rev.}
  {\bf D77} (2008) 094502, [\href{http://xxx.lanl.gov/abs/0705.4295}{{\tt
  arXiv:0705.4295}}].

\bibitem{Alexandrou:2013cda}
C.~Alexandrou, M.~Constantinou, V.~Drach, K.~Hatziyiannakou, K.~Jansen, {\em
  et.~al.}, {\it {Nucleon Structure using lattice QCD}},  {\em Nuovo Cim.} {\bf
  C036} (2013), no.~05 111--120, [\href{http://xxx.lanl.gov/abs/1303.6818}{{\tt
  arXiv:1303.6818}}].

\bibitem{Constantinou:2014tga}
M.~Constantinou, {\it {Hadron Structure}},  {\em PoS} {\bf LATTICE2014} (2014)
  001, [\href{http://xxx.lanl.gov/abs/1411.0078}{{\tt arXiv:1411.0078}}].

\bibitem{Alekhin:2013nda}
S.~Alekhin, J.~Bluemlein, and S.~Moch, {\it {The ABM parton distributions tuned
  to LHC data}},  {\em Phys.Rev.} {\bf D89} (2014) 054028,
  [\href{http://xxx.lanl.gov/abs/1310.3059}{{\tt arXiv:1310.3059}}].

\bibitem{Sergey}
S.~Alekhin, {\it {Private communication}}, .

\bibitem{Alexandrou:2013wca}
C.~Alexandrou, M.~Constantinou, V.~Drach, K.~Hadjiyiannakou, K.~Jansen, {\em
  et.~al.}, {\it {Evaluation of disconnected quark loops for hadron structure
  using GPUs}},  {\em Comput.Phys.Commun.} {\bf 185} (2014) 1370--1382,
  [\href{http://xxx.lanl.gov/abs/1309.2256}{{\tt arXiv:1309.2256}}].

\bibitem{Alexandrou:2013nda}
C.~Alexandrou, M.~Constantinou, S.~Dinter, V.~Drach, K.~Hadjiyiannakou, {\em
  et.~al.}, {\it {Strangeness of the nucleon from Lattice Quantum
  Chromodynamics}},  \href{http://xxx.lanl.gov/abs/1309.7768}{{\tt
  arXiv:1309.7768}}.

\bibitem{Abdel-Rehim:2013wlz}
A.~Abdel-Rehim, C.~Alexandrou, M.~Constantinou, V.~Drach, K.~Hadjiyiannakou,
  {\em et.~al.}, {\it {Disconnected quark loop contributions to nucleon
  observables in lattice QCD}},  {\em Phys.Rev.} {\bf D89} (2014), no.~3
  034501, [\href{http://xxx.lanl.gov/abs/1310.6339}{{\tt arXiv:1310.6339}}].

\bibitem{Iwasaki:1985we}
Y.~Iwasaki, {\it {Renormalization Group Analysis of Lattice Theories and
  Improved Lattice Action: Two-Dimensional Nonlinear O(N) Sigma Model}},  {\em
  Nucl. Phys.} {\bf B258} (1985) 141--156.

\bibitem{Frezzotti:2000nk}
{\bf Alpha} Collaboration, R.~Frezzotti, P.~A. Grassi, S.~Sint, and P.~Weisz,
  {\it {Lattice QCD with a chirally twisted mass term}},  {\em JHEP} {\bf 0108}
  (2001) 058, [\href{http://xxx.lanl.gov/abs/hep-lat/0101001}{{\tt
  hep-lat/0101001}}].

\bibitem{Frezzotti:2003ni}
R.~Frezzotti and G.~C. Rossi, {\it {Chirally improving Wilson fermions. I: O(a)
  improvement}},  {\em JHEP} {\bf 08} (2004) 007,
  [\href{http://xxx.lanl.gov/abs/hep-lat/0306014}{{\tt hep-lat/0306014}}].

\bibitem{Frezzotti:2003xj}
R.~Frezzotti and G.~C. Rossi, {\it Twisted-mass lattice qcd with mass
  non-degenerate quarks},  {\em Nucl. Phys. Proc. Suppl.} {\bf 128} (2004)
  193--202, [\href{http://xxx.lanl.gov/abs/hep-lat/0311008}{{\tt
  hep-lat/0311008}}].

\bibitem{Frezzotti:2004wz}
R.~Frezzotti and G.~C. Rossi, {\it Chirally improving wilson fermions. ii:
  Four-quark operators},  {\em JHEP} {\bf 10} (2004) 070,
  [\href{http://xxx.lanl.gov/abs/hep-lat/0407002}{{\tt hep-lat/0407002}}].

\bibitem{Baron:2010bv}
R.~Baron {\em et.~al.}, {\it {Light hadrons from lattice QCD with light (u,d),
  strange and charm dynamical quarks}},  {\em JHEP} {\bf 06} (2010) 111,
  [\href{http://xxx.lanl.gov/abs/1004.5284}{{\tt arXiv:1004.5284}}].

\bibitem{Baron:2010th}
{\bf European Twisted Mass} Collaboration, R.~Baron {\em et.~al.}, {\it
  {Computing K and D meson masses with $N_f = 2+1+1$ twisted mass lattice
  QCD}},  \href{http://xxx.lanl.gov/abs/1005.2042}{{\tt arXiv:1005.2042}}.

\bibitem{Alexandrou:2008tn}
{\bf European Twisted Mass} Collaboration, C.~Alexandrou {\em et.~al.}, {\it
  {Light baryon masses with dynamical twisted mass fermions}},  {\em Phys.
  Rev.} {\bf D78} (2008) 014509, [\href{http://xxx.lanl.gov/abs/0803.3190}{{\tt
  arXiv:0803.3190}}].

\bibitem{Michael:2007vn}
{\bf ETM Collaboration} Collaboration, C.~Michael and C.~Urbach, {\it {Neutral
  mesons and disconnected diagrams in Twisted Mass QCD}},  {\em PoS} {\bf
  LAT2007} (2007) 122, [\href{http://xxx.lanl.gov/abs/0709.4564}{{\tt
  arXiv:0709.4564}}].

\bibitem{Boucaud:2008xu}
{\bf ETM} Collaboration, P.~Boucaud {\em et.~al.}, {\it {Dynamical Twisted Mass
  Fermions with Light Quarks: Simulation and Analysis Details}},  {\em Comput.
  Phys. Commun.} {\bf 179} (2008) 695--715,
  [\href{http://xxx.lanl.gov/abs/0803.0224}{{\tt arXiv:0803.0224}}].

\bibitem{Alexandrou:2010me}
C.~Alexandrou, M.~Constantinou, T.~Korzec, H.~Panagopoulos, and F.~Stylianou,
  {\it {Renormalization constants for 2-twist operators in twisted mass QCD}},
  {\em Phys.Rev.} {\bf D83} (2011) 014503,
  [\href{http://xxx.lanl.gov/abs/1006.1920}{{\tt arXiv:1006.1920}}].

\bibitem{Blossier:2014kta}
{\bf ETM Collaboration} Collaboration, B.~Blossier {\em et.~al.}, {\it
  {Renormalization of quark propagator, vertex functions and twist-2 operators
  from twisted-mass lattice QCD at $N_f$=4}},
  \href{http://xxx.lanl.gov/abs/1411.1109}{{\tt arXiv:1411.1109}}.

\bibitem{Blossier:2014rga}
B.~Blossier, M.~Brinet, P.~Guichon, V.~Morénas, O.~Pène, {\em et.~al.}, {\it
  {Renormalization constants for $N_{\rm f}=2+1+1$ twisted mass QCD}},
  \href{http://xxx.lanl.gov/abs/1411.0053}{{\tt arXiv:1411.0053}}.

\bibitem{Alexandrou:2014sha}
C.~Alexandrou, V.~Drach, K.~Jansen, C.~Kallidonis, and G.~Koutsou, {\it {Baryon
  spectrum with $N_f=2+1+1$ twisted mass fermions}},
  \href{http://xxx.lanl.gov/abs/1406.4310}{{\tt arXiv:1406.4310}}.

\bibitem{Dinter:2011sg}
S.~Dinter, C.~Alexandrou, M.~Constantinou, V.~Drach, K.~Jansen, {\em et.~al.},
  {\it {Precision Study of Excited State Effects in Nucleon Matrix Elements}},
  {\em Phys.Lett.} {\bf B704} (2011) 89--93,
  [\href{http://xxx.lanl.gov/abs/1108.1076}{{\tt arXiv:1108.1076}}].

\bibitem{Abdel-Rehim:2013yaa}
A.~Abdel-Rehim, P.~Boucaud, N.~Carrasco, A.~Deuzeman, P.~Dimopoulos, {\em
  et.~al.}, {\it {A first look at maximally twisted mass lattice QCD
  calculations at the physical point}},  {\em PoS} {\bf LATTICE2013} (2014)
  264, [\href{http://xxx.lanl.gov/abs/1311.4522}{{\tt arXiv:1311.4522}}].

\bibitem{Abdel-Rehim:2014nka}
A.~Abdel-Rehim, C.~Alexandrou, P.~Dimopoulos, R.~Frezzotti, K.~Jansen, {\em
  et.~al.}, {\it {Progress in Simulations with Twisted Mass Fermions at the
  Physical Point}},  {\em PoS} {\bf LATTICE2014} (2014) 119,
  [\href{http://xxx.lanl.gov/abs/1411.6842}{{\tt arXiv:1411.6842}}].

\end{thebibliography}\endgroup

\end{document}